# CMS Conider Report

# CMS Conference Report

Mailing address CMS CERN, CH-1211 GENEVA 23, Switzerland

August 15, 2002

# Database independent Migration of Objects into an Object-Relational Database

Kamran Munir [1], M. Waseem Hassan [1, 2], Arshad Ali [1], R. McClatchey [2] and I. Willers [3]

1- National University of Science & Technology (NUST)
Tamiz Uddin Road, P.O.Box 297, Rawalpindi, Pakistan

2- Centre for Complex Cooperative Systems (CCCS)
UWE, Bristol BS16 1QY UK

3- European Organization for Nuclear Research (CERN)
1211 Geneva, Switzerland

**Abstract**

CERN's (European Organization for Nuclear Research) WISDOM project [1] deals with the replication of data between homogeneous sources in a Wide Area Network (WAN) using the extensible Markup Language (XML). The last phase of the WISDOM (Wide-area, database Independent Serialization of Distributed Objects for data Migration) project [2], indicates the future directions for this work to be to incorporate heterogeneous sources as compared to homogeneous sources as described by [3]. This work will become essential for the CERN community once the need to transfer their legacy data to some other source, other then Objectivity [4], arises. Oracle 9i – an Object-Relational Database (including support for abstract data types, ADTs) appears to be a potential candidate for the physics event store in the CERN CMS experiment as suggested by [4] & [5]. Consequently this database has been selected for study. As a result of this work the HEP community will get a tool for migrating their data from Objectivity to Oracle9i.

# 1. Introduction

The Compact Muon Solenoid experiment, CMS [1] has a large number of Terabyte sized databases and the experiment is due to take its first data in mid 2007 after which a PetaByte of raw event data will be generated and stored per year. Until very recently the object-oriented database, Objectivity/DB was the first database of choice for the CMS Event Store. But In July 2001, CMS decided [4] to evaluate the Oracle9i database as a potential candidate for its baseline for persistent data storage. This decision was mainly motivated by concerns about the trends concerning the market performance of the Objectivity/DB, which was the baseline of CMS at that time. So far there is no indication about the final decision of the next frontline database of choice in CMS, but there seems to be an inclination in the HEP community towards using Oracle9i as part of the next possible solution [5].

If for example the HEP community select Oracle9i or any Object-Relational Database for the event store or its metadata then there will arise a need to transfer legacy data from object-oriented databases to object-relational databases. In this context, there is a need to find a solution for this activity.

CERN's WISDOM project [1] deals with the replication of data in a Wide Area Network (WAN) in a database independent format i.e. the widely acceptable standard for data exchange – the extensible Markup Language (XML) [6]. The WISDOM project, according to [2], [3] & [7], provides tools for converting object-oriented data into XML and back i.e. conversion of the XML objects into the database (Objectivity/DB). This means that objects are transferred from one layer of persistence (i.e. OODB) to another layer of persistence (i.e. XML). But the difference between the two is that the second form has an added feature in addition to persistency i.e. mobility.

The XML generated as a result of serialization [2], can be directly used for migrating objects to yet another layer of persistence, e.g. Object-Relational Databases. Here it will also be appropriate to highlight one of the features of Oracle9i concerning its support for XML. The Oracle8i/9i understands XML data, can store XML documents and also provides a query facility for the XML data. In addition to this Oracle9i has also made a breakthrough by facilitating the much-awaited ADT support.

In the light of the above, research has been proposed, and then conducted to exploit the capabilities of the tools developed in the WISDOM project and the current rich support provided by Oracle9i, to design a mechanism to transport data in a heterogeneous environment consisting of both object-oriented and object-relational data. The medium for transportation is obviously the database independent XML. This research work can also be referred to as an extension to the work being done in the context of the WISDOM project. Since the WISDOM project is a sister project to the CRISTAL project [8], so the results of this research can also be exploited in the CRISTAL project. Keeping this in mind for real database tests, the CRISTAL databases have been used.

In this paper a detailed study will be reported on the following core issues:

- ❑ Transformation of (all or part of) the data from Oracle database tables or views into XML.

- ❑ Extraction of data from an XML document by using canonical mapping, and insertion of data into the appropriate columns/attributes of a table in Oracle Database.

- ❑ Mapping of objects into relational tables

- ❑ Analyzing the structure information of the objects residing in the Objectivity database and extracting the schema definition to convert it in a format suitable for designing schema in an RDBMS e.g. Oracle.

- ❑ Extraction of objects from the object XML file generated from Objectivity database and to migrate them into the Oracle database by making use of the map file, see figure 1.

The set of tools developed as a result of this research are the deliverables for this project. Currently these tools have been tested with the CRISTAL database.

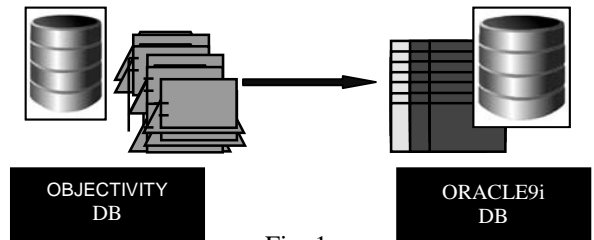

Fig. 1

**Next sections include:**

- ❑ Mapping OODB schema into RDBMS
- ❑ General factors while mapping objects
- ❑ Issues related to storing data from XML documents to traditional databases.
- ❑ Survey of XML database products and XML databases
- ❑ Oracle XML components
- ❑ Limitations while using Oracle XML component (e.g. XSU)
- ❑ The project scenario
- ❑ Transformation of data between XML documents generated by objectivity and relational databases e.g. Oracle.
- ❑ Prototype for transfer XML objects in Oracle database
- ❑ Future directions and conclusion

## 2. Related work

XML is becoming the Internet standard for information exchange. Businesses need to be able to communicate with other businesses and workflow components by using XML. A number of projects have mapped XML Schemas to object classes. Currently there is much effort going on in using XML as a means of serializing objects. The following research areas are distinguished: serialisation and deserialisation between XML and an object-relational database, serialisation and deserialisation of pure objects into an object-relational database and migration of objects between commercial databases using XML [14].

### 2.1 Serialisation and deserialisation of XML into object-relational DB:

Object-relational technology provides different methods and tools for importing and exporting XML information to and from a database e.g. XSU and XSQL. A number of third party tools are also available to support this feature, the most notable being XML_RDB, XMLServlet and XMLShark. A detailed analysis of XSU and XSQL has been carried out.

The XML-SQL Utility (XSU) [9] transforms XML into Oracle tables and vice versa. This work is relevant to the current work in the context that it addresses the transformation of XML information into Oracle tables but in this case the structure of the resultant XML document is fixed i.e. it conforms to a particular DTD agreed by the designers of the product. Moreover by using this utility, the schema cannot be created dynamically. This contrasts with the current work, which addresses the conversion of XML objects from an object source into tables inside Oracle.

### 2.2 Serialisation and deserialisation of pure objects (XML) into object-relational DB:

In order to transform objects embedded in an XML document into tables, there is a need to map core object-oriented concepts like inheritance, polymorphism etc. into tables. For this purpose different third party tools were investigated. Out of the list of these tools XML_DBMS was the most relevant to the current research. The functionalities of this tool include extended functionalities which make it one of the most powerful.

## 2.3 Migration of objects between commercial databases using XML:

Currently there is no work being done in the context of migrating objects (after converting them into XML) between different commercially available pure object-oriented and object-relational products. The research work reported in this paper address this deficiency by providing a tool, which can convert XML information (produced by using tools from [2], [3]) from an object-oriented database (i.e. Objectivity/DB) to object-relational database (i.e. Oracle).

# 3. Object Database vs. Object-Relational Databases.

We need to understand the differences between object database management systems and the object relational database management systems (ORDBMSs).
Relational databases (RDBs) are far more common than OODBs. Relational databases store information in tables; a table consists of any number of rows, each row containing several columns of information. (Rows are more formally called relations, which is where the term ''relational database'' originates.)

Object databases employ a data model that has object-oriented aspects like class, with attributes and methods and integrity constraints; provides object identifiers (OIDs) for any persistent instance of a class; supports encapsulation (data and methods); multiple inheritance; and supports abstract data types. Object databases combine the elements of object orientation and object-oriented programming languages with database capabilities. They provide more than persistent storage of programming language objects. Object databases extend the functionality of object programming languages (e.g., C++, Smalltalk, Java) to provide full-featured database programming capability.

## 3.1 Architecture Differences between RDB and ODB

Primarily, RDBMSs have been built around central server architectures, which are much the same as mainframe architectures. ODBMSs often assume a network of computers, with processing on the back or front end, as well as intermediate tiers, caching on each level, and clustering capabilities independent of type. In terms of computation model, although RDBMSs typically confine all processing to the SQL language and its operations (SELECT/PROJECT/JOIN and INSERT/UPDATE/DELETE), ODBMSs allow the use of host object languages like C++ and Java directly on the objects "in the database"; that is, instead of translating back and forth between application language structures (COBOL, C, etc.) and database structures (SQL), application programmers can simply use the object language to create and access objects through the methods. The database system maintains the persistence, integrity, recoverability, and concurrency of those same objects.

## 3.2 Database Management System Products by Vendor:

| Vendor | ORDBMS | Vendor ODBMS |
|--------|--------|--------------|
| Oracle | Oracle 8.x, 9.x | |
| Informix | Universal Server | |
| IBM | Universal Database (DB/2 Extenders) | Computer Associates Jasmine |
| UniSQL | UniSQL/X | |
| Unisys | OSMOS | Gemstone |
| | | Gemstone |

## 3.3 Comparison of Database Management Systems:

| Criteria | ORDBMS | ODBMS |
|---|---|---|
| Defining standard | SQL3/4 | ODMG-V2.0 |
| Support for object-oriented programming | Limited mostly to new data types | Direct and extensive |
| Simplicity of use | Same as RDBMS, with some confusing extensions | OK for programmers; some SQL access for end users |
| Simplicity of development | Provides independence of data from application, good for simple relationships | Objects are a natural way to model; can accommodate a wide variety of types and relationships |
| Extensibility and content | Limited mostly to new data types | Can handle arbitrary complexity; users can write methods and on any structure |
| Complex data relationships | Required much expertise to model | Can handle arbitrary complexity; users can write methods and on any structure |
| Performance versus interoperability | Level of safety varies with vendor, must be traded off; achieving both requires extensive testing | Level of safety varies with vendor; most ODBMSs allow programmers to extend DBMS functionality by defining new classes |
| Distribution, replication, and federated databases | Extensive | Varies with vendor; a few provide extensive support |
| Product maturity | Immature; extensions are new, are still being defined, and are relatively unproven | Relatively mature |
| Legacy people and the universality of SQL | Can take advantages of RDBMS tools and developers | SQL accommodated, but intended for object-oriented programmers. |
| Software ecosystems | Provided by major RDBMS companies | ODBMS vendors beginning to emulate RDBMS vendors, but none offers large markets to other ISVs |
| Vendor viability | Expected for the major RDBMS vendors; UniSQL is struggling | Less of an issue than it was; some shakeout still expected |

Source: International Data Corporation

| | Strengths | Weaknesses |
|---|---|---|
| **OODB** | • Good abstraction and modeling capabilities<br>• Seamless integration of Java and database objects<br>• Better performance for some applications<br>• Java Report (OStore, Poet, Oracle) | • Good Java language skills needed up-front<br>• Total market share still small<br>• Long-term survival/commitment of some vendors unknown<br>• Many users moving straight to Object-Relational<br>• Less number of professionals are available then relational database experts |
| **ORDB** | • Better integration with Java than Relational<br>• Provides smoother upgrade path for heavy Relational users<br>• Provides fall-back to just Relational<br>• Leverage existing SQL skills, | • Proprietary SQLJ extensions<br>• Complex structures still flattened for storage<br>• Generating bulk-load file for complex data could be difficult |

| | investment |

The ORDBMS vendors are much larger and have huge entrenched marketing infrastructure. By comparison, the ODBMS vendors are much smaller.

# 4. Object-Relational Mapping

Mapping Objects to tables is a problem that has been around as long as there has been a need to program in an object-oriented language. However, there are increasingly compelling reasons to prefer relational databases (or object-relational databases) over object-oriented database for data persistence. These reasons include better integration with Java, provision of a smoother upgrade path for legacy relational database users, provision of a fall-back philosophy to another relational database, leveraging existing SQL skills, investment from major players in the corporate sector etc. In essence, there is a requirement to handle the mapping of objects into relational tables and to bridge the gap between the two different persistency options i.e. OODBMS and RDBMS.

## 4.1 Mapping OODB schema into RDBMS

In order to model the concepts of object orientation into relational table structures, there is a need to handle some mapping issues like ***aggregation, inheritance, polymorphism*** and ***associations*** between classes while migrating towards RDBMS from OODB. These issues are covered in the table below:

| Single Table Aggregation | Map aggregation to a relational data model by integrating all aggregated object's attributes into a single table. Then the aggregated object's attributes into the same table as the aggregating object's [16]. |
|---|---|
| Foreign Key Aggregation | Foreign Key Aggregation is the usual way to map 1:n associations. A separate table is used for the aggregated type. The Aggregating Object is mapped to a table. The Aggregated Object is mapped to another table. |
| One Inheritance Tree One Table: | The union of all attributes of all objects in the inheritance hierarchy will be used as the columns of a single database table. And Null values to fill the unused fields in each record [15]. |
| One Class One Table: | The attributes of each class are mapped to a separate table. A synthetic OID will be inserted into each table to link derived classes rows with their parent table's corresponding rows [13]. |
| Association Table [n:m associations]: | A separate table will be created containing the Object Identifiers (or Foreign Keys) of the two object types participating in the association. Map the rest of the two object types to tables using any other suitable mapping patterns presented in paper [13] [15]. |

It should be noted that an abstract class is also mapped to a separate table.

## 4.2 General factors while mapping objects

The major factors that should be taken into account while mapping objects to tables are:

| Performance: | The way objects are mapped to tables has significant influence on the number of database accesses that occur in a system. It is therefore a good idea to waste a few processor cycles and some RAM memory to economize on slow IO [14][16]. |
|---|---|
| Read versus write/update performance | To be sure about the frequency of read and write/update operations before finalizing a certain table design. |
| Flexibility and maintenance cost | Flexibility is more important than performance is schema evolution as attributes will be often deleted or inserted, or classes added or deleted and class hierarchies restructured. Once the hierarchy and classes become stable then it may be desirable to switch to a mapping with optimal performance. |
| Space consumption versus Performance: | There are mappings that use no surplus database space (e.g. fields with null values) and others that leave large portions of a database record unused [16]. |
| Query processing: | There are two conflicting purposes; Firstly, data have to serve in an |

| | information system. Secondly, data have to be ready for online transaction processing with good performance. |
|---|---|

Finally, we should not forget that objects consist of methods as well as data. Very few databases have ever offered the facility to store both methods and data.

# 5. Survey of XML Database Products

There are various ways to solve the problem of effective, automatic conversion of XML data into and out of relational databases. Database vendors such as IBM, Microsoft, Oracle, and Sybase have developed tools to assist in converting XML documents into relational tables.

XML documents fall into two broad categories: *data-centric* and *document-centric*. **Data-centric** documents are those where XML is used as a data transport.

For data-centric applications, the following is the list of **Commercial XML products** that support both transfers database to XML and XML to database:

| Product | Developer |
|---|---|
| Attunity Connect | Attunity Ltd. |
| XML Servlet | Cerium Component |
| XChainJ | Cogent Logic Corp. |
| TransVerse | Coyote Consultants |
| XML Junction, Data Junction | Data Junction, Inc. |
| jXTransformer | DataDirect Technologies |
| Import/Export Studio | Etasoft |
| Allora | HiT Software |
| XMLShark | infoShark |

Oracle has provided some **Development-only products** like XSQL Servlet and XML SQL Utility, where XML SQL Utility supports both database to XML and XML to database and XSQL Servlet supports only database to XML transfers.

Following are the **Open source XML products** that support both conversions from database to XML and XML to database transfers by the use of mapping. And they support the Rational Database Type.

| Product | Developer |
|---|---|
| JaxMe | Jochen Wiedmann |
| DBIx::XML_RDB | Matt Sergeant |
| XML-DBMS | Ronald Bourret, et al |

## 5.1 XML Enabled Databases

To store and retrieve the data in data-centric documents, an XML-enabled database is required that is tuned for data storage, similar to a relational or object-oriented database, and some sort of data transfer software which might be built into the database (in this case the database is said to be XML-enabled) or any third-party middleware can be used. The following is a list of the Relational databases that are XML enabled and having commercial License [15].

| Product | Developer |
|---|---|
| Access 2002 | Microsoft |
| DB2 XML Extender, DB2 Text Extender | IBM |
| Informix | IBM |
| Microsoft SQL Server 2000 | Microsoft |

Objectivity/DB (from Objectivity, Inc.) is an Object Oriented database, which has recently announced its XML interface.

# 6. Middleware Oracle XML components

In this case middleware, by definition is the software used by the data-centric applications to transfer data between XML documents and databases. These are written in a variety of languages and can be used with any of these database drivers e.g. ODBC, JDBC, or OLE DB. Most of these middleware components can send data across the Internet but if there is a need to access data from remote locations then these components should be used integrated within a Web server. Middleware products range from homegrown projects to commercial data conversion engines.

## 6.1 Oracle XML Components

There follows an overview of Oracle's XML components that can be used to transform XML data to and from Oracle. After that the limitations of using these components while migrating Objects from Objectivity/DB into Oracle will be elaborated.

In the newest releases of Oracle e.g. Oracle9*i* - several components, utilities, and interfaces taking advantage of XML technology in building Web-based database applications, are provided. The selection of a set of components for any required scenario depends on application requirements, programming preferences, development and deployment environment. The following XML components are provided with Oracle9*i* and Oracle9*i* Application Server:

### 6.1.1 XML Developer's Kit (for java)

XDK for Java is composed of the following components:

❑ **XML Parser for Java:** It creates and parses XML by using industry standard DOM and SAX interfaces. It also includes an XSL Transformation (XSLT) processor that transforms XML to XML or other text-based formats, such as HTML.
❑ **XML Schema Processor for Java:** It supports simple and complex types and is built on top of the Oracle XML Parser for Java version 2.
❑ **XML Class Generator for Java:** It creates source files from an XML DTD or XML Schema definition.
❑ **XSQL Servlet:** XSQL Servlet processes SQL queries embedded in an XSQL file e.g. "xxx.xsql" which returns results in the XML format by using the XML SQL Utility and XML Parser for Java.
❑ **XML SQL Utility (XSU) for Java:** XSU enables us to transform data retrieved from object-relational database tables or views into XML and similarly to extract data from an XML document (details given in section 5.1.2).
❑ **Storage XML documents in Character Large Objects, CLOBs:** If the incoming XML documents do not conform to one particular structure, then it might be better to store such documents in CLOBs.

### 6.1.2 XML-SQL Utility (XSU):

XML has rapidly become the format for data interchange; at the same time, a substantial amount of business data resides in object-relational databases. It is therefore necessary to have the ability to transform this "relational" data into XML.

The XML-SQL Utility (XSU) enables us to do the following things:

❑ It can transform data retrieved from object-relational database tables or views into XML.
❑ By using XSU, data can be extracted from an XML document and by using a canonical mapping this data can be inserted into the appropriate columns/attributes of a table or a view.
❑ By using XSU, data can be extracted from a XML document and this data can be applied for updating or deleting values of the appropriate columns/attributes.

XSU is composed of Java classes and these Java classes can be loaded into a Java-enabled Oracle8i/9i database; furthermore, the XSU contains a PL/SQL wrapper that publishes the XSU's Java API to PL/SQL by creating a PL/SQL API. In this way new Java applications can be written that run inside the database and which directly access the XSU's Java API; in addition to this functionality PL/SQL applications can also be written that access XSU through its PL/SQL API the XSU's functionality can be accessed directly through SQL. It is noted that to load and run Java code inside the database we need a Java-enabled Oracle8i (or later) Server.

The Java programs make use of the XSU through its Java API for the purpose of XML generation in the and for integration with different JDBC data sources.

For example, if the query "select * from emp" is specified, the XSU would generate the following XML document

```
CREATE TABLE emp
(
  EMPNO NUMBER,
  ENAME VARCHAR2(20),
  JOB VARCHAR2(20),
  MGR  NUMBER,
  HIREDATE DATE,
  SAL NUMBER,
  DEPTNO NUMBER
);
```

Fig. 2(a)

```
<?xml version='1.0'?>
<ROWSET>
<ROW num="1">
<EMPNO>7369</EMPNO>
<ENAME>Smith</ENAME>
<JOB>CLERK</JOB>
 <MGR>7902</MGR>
 <HIREDATE>12/17/1980 0:0:0</HIREDATE>
<SAL>800</SAL>
<DEPTNO>20</DEPTNO>
 </ROW>
 <!-- additional rows ... -->
</ROWSET>
```

Fig. 2(b)

In the generated XML, the rows returned by the SQL query are enclosed in a *ROWSET* tag (Fig. 2(b)) to make the *<ROWSET>* element, which is also the root element of the generated XML document. The *<ROWSET>* element contains one or more *<ROW>* elements. Each of these *<ROW>* elements contains the data from one of the returned database rows.

### 6.2 Limitations while using XSU

The current study has revealed the following limitations in using XSU:

❑   Currently the XML-SQL Utility (XSU) can only store to a single table [12].

❑   Due to a number of limitations of the DTD, XML SQL Utility cannot generate the database schema with DTD [12].

❑   If XML files need to be stored as CLOBs in the Oracle database, the maximum file size that can be stored is 2 GB [12].

# 7. Other Issues

This section discusses a number of issues related to storing data from XML documents to traditional databases. Generally, there is no choice about how data transfer software resolves these issues. However, it should be noted that these issues exist, as they might help in the selection of the correct software. Some of these issues are discussed in the following sections.

## 7.1 Data Types

XML does not support data types in any meaningful sense. Except for unparsed entities, all data in an XML document is text, even if it represents another data type, such as a date or an integer.

## 7.2 Null Data

 In the database world, null data means data that simply isn't there. This is very different from a value of 0 (for numbers) or zero length (for a string). For example, if data has been collected from a weather station and if the thermometer is not working, then a null value would be stored in the database rather than a 0, which would mean something different altogether.

## 7.3 Normalization

*Normalization* refers to the process of designing a database schema in which a given piece of data is represented only once. Normalization has several obvious advantages, such as reducing disk space and eliminating the possibility of inconsistent data, which can occur when a given piece of data is stored in more than one place. It is one of the cornerstones of relational technology and is a flashpoint in discussions about storing data in native XML databases.

# 8. The Project Scenario

The methodology that is being followed in the current research in order to transfer data from Objectivity to Oracle is presented in the figure below (Fig. 3).

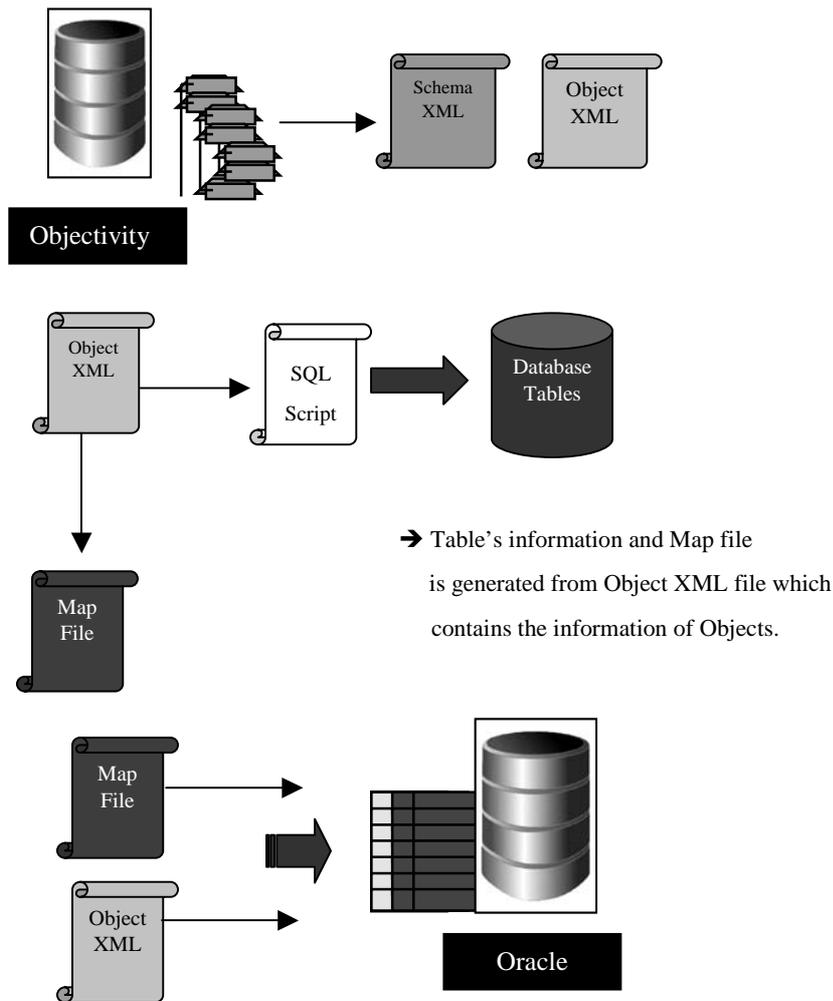

→ Table's information and Map file is generated from Object XML file which contains the information of Objects.

Fig. 3

As noted earlier data from Objectivity can be serialized in XML format by tools provided by [2,3]. The produced XML (i.e. objects) include information from the schema and the data from the Objectivity Database, such that two XML files are generated [2] for a single database. Out of these, the Schema XML file contains information about the schema defined in the Objectivity database for object storage. The second one is the Object XML file that contains the tangible data from Objectivity/DB.

In order to design the schema in a relational database, the corresponding schema information is required from Objectivity. That schema information can be taken from the Object-XML file (generated from the Objectivity database).
Here two files are required for further processing, namely:

   a.  SQL script file: It contains information about the Oracle schema
   b.  Map file: It contains information about objects which are to be mapped into the Oracle database

By using the SQL script file tables can be created in database. After creating tables (by using ODBC connection), "ObjXMLfile" (which has tangible data of objectivity in XML format) is then mapped into database by using map file (generated from a XML file).

## 8.1. DTD used for Serialization (by WISDOM project [1-3])

The following is the sample DTD proposed by the WISDOM project. (See Fig. 4)

**Hierarchical model of the schema structure**

The following (Fig. 5) is the model of the schema structure extracted from the DTD shown in Fig. 4.

```
<?xml version='1.0' encoding='utf-8' ?>
<!DOCTYPE Schema
[<!ELEMENT Schema (TopLevelModule)>
<!ELEMENT TopLevelModule (Module*,Class*)>
<!ELEMENT Module (Module*,Class*)>
<!ELEMENT Class (BaseClass*,Attributes?,Relationships?)>
<!ELEMENT BaseClass EMPTY>
<!ELEMENT Attributes (BasicAttribute*,RefAttribute*,
EmbeddedClassAttribute*,VArrayBasicAttribute*,
VArrayEmbeddedClassAttribute*,VArrayRefAttribute*)>
<!ELEMENT BasicAttribute EMPTY>
<!ELEMENT RefAttribute EMPTY>
<!ELEMENT EmbeddedClassAttribute EMPTY>
<!ELEMENT VArrayBasicAttribute EMPTY>
<!ELEMENT VArrayRefAttribute EMPTY>
<!ELEMENT VArrayEmbeddedClassAttribute EMPTY>
<!ELEMENT Relationships (Unidirectional*,Bidirectional*)>
<!ELEMENT Unidirectional EMPTY>
<!ELEMENT Bidirectional EMPTY>

<!--Attribute Definitions -->
<!ATTLIST Module SchemaNumber CDATA #REQUIRED>
<!ATTLIST Module Name CDATA #REQUIRED>
(Contd….)
```

Fig. 4

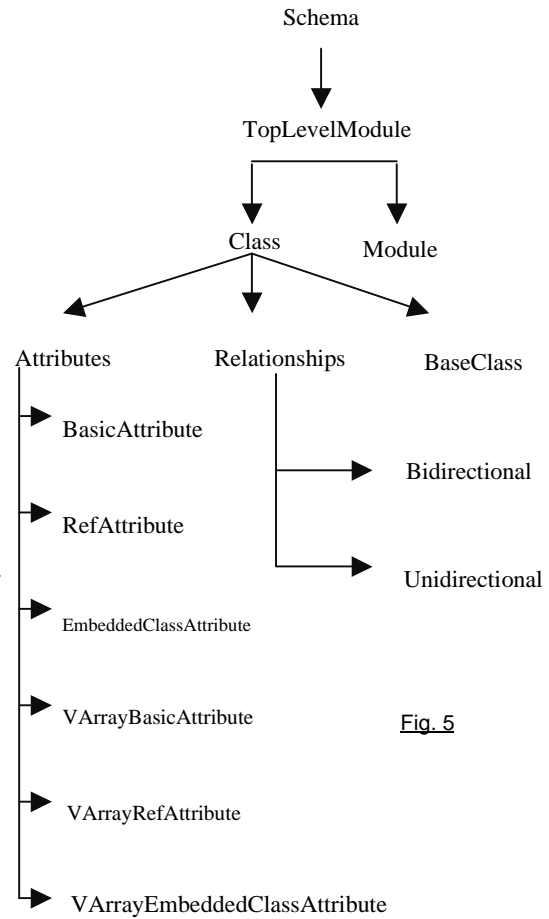

Fig. 5

## 8.2 Object-relational mapping

Object-relational mapping is used to map objects to the relational database. For example, firstly consider the simple Fig. 6(a) XML document:

```
<SalesOrder>
    <Number>1234</Number>
    <Customer>Gallagher Industries</Customer>
    <Date>29.10.00</Date>
    <Line Number="1">
      <Part>A-10</Part>
      <Quantity>12</Quantity>
      <Price>10.95</Price>
    </Line>
    <Line Number="2">
      <Part>B-43</Part>
      <Quantity>600</Quantity>
      <Price>3.99</Price>
    </Line>
</SalesOrder>
```

Fig. 6(a)

In Fig. 6(a) there is a simple XML document, when this document is mapped in relational table structure then the schema where this information (data in XML file) will be stored is shown in Fig. 6(b). In Fig. 7(a), 7(b) actual tables are described with data, rows in these tables are linked through a primary key or foreign key relationship.

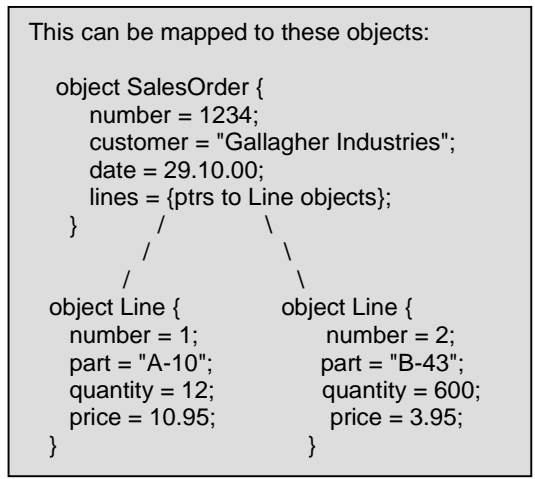

Fig. 6(b)

**Table SaleOrders**

| Number | Customer | Date |
|--------|----------|------|
| 1234 | Gallagher Industries | 29.10.00 |

Fig. 7(a)

**Table Lines**

| SONumber | Line | Part | Quantity | Price |
|----------|------|------|----------|-------|
| 1234 | 1 | A-10 | 12 | 10.95 |
| 1234 | 2 | B-43 | 600 | 3.99 |
| ... | ... | ... | ... | ... |

Fig. 7(b)

## 8.3 Mapping of CRISTAL database schema in Relational database

When the hierarchical model of the schema (see Fig. 5) is mapped to a relational database, the ER design that is produced is shown in the following figures i.e. Fig. 8. The relationships here are described with primary and foreign keys, e.g. as there can be multiple modules in a single schema (see Fig. 4) so TOPLEVELMODULE has the master detail relationship with the schema table. Similarly CLASS and MODULE has the master detail relationship with TOPLEVELMODULE and so on.

<!ELEMENT Schema (TopLevelModule)>

<!ELEMENT TopLevelModule (Module*,Class*)>

<!ELEMENT Module (Module*,Class*)>

<!ELEMENT Class (BaseClass*,Attributes?,Relationships?)>

<!ELEMENT BaseClass EMPTY>

<!ELEMENT Attributes
(BasicAttribute*,RefAttribute*,EmbeddedClassAttribute*,VArrayBasicAttribute*,
VArrayEmbeddedClassAttribute*,VArrayRefAttribute*)>

<!ELEMENT BasicAttribute EMPTY>

<!ELEMENT RefAttribute EMPTY>

<!ELEMENT EmbeddedClassAttribute EMPTY>

<!ELEMENT VArrayBasicAttribute EMPTY>

<!ELEMENT VArrayRefAttribute EMPTY>

<!ELEMENT VArrayEmbeddedClassAttribute EMPTY>

<!ELEMENT Relationships (Unidirectional*,Bidirectional*)>

<!ELEMENT Unidirectional EMPTY>

<!ELEMENT Bidirectional

<!ATTLIST Module SchemaNumber CDATA #REQUIRED>

<!ATTLIST Module Name CDATA #REQUIRED>

<!ATTLIST Class Name CDATA #REQUIRED>

…… …… …… ……

…… …… …… ……

CRISTAL Schema in Relational Table Structure

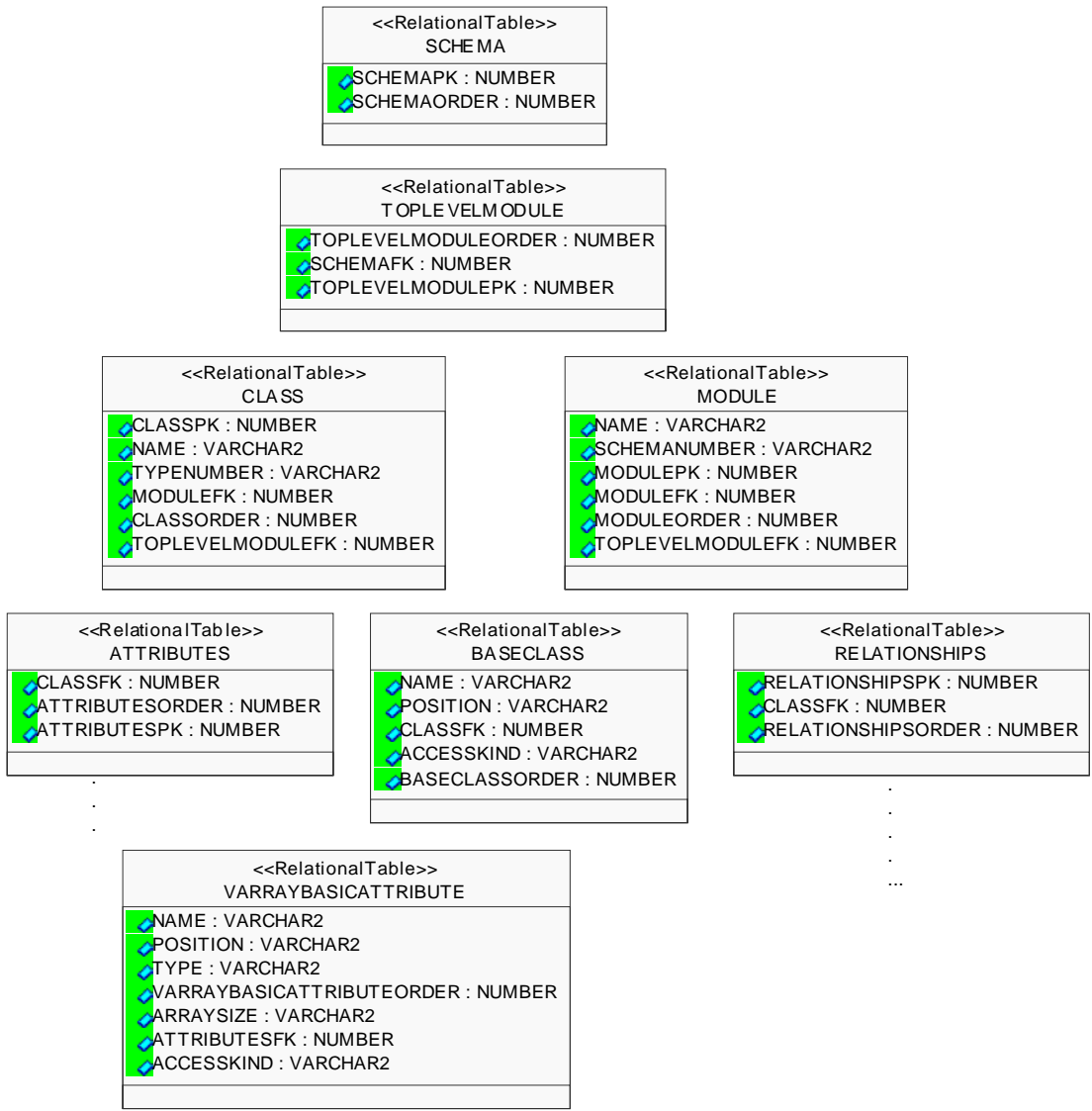

## 9. The transformation of data from XML documents to relational DB

There are several freeware XML products (e.g. JaxMe, XML_RDB and XML-DBMS etc) that can be used for XML data transfer into a relational database. Out of these, XML-DBMS has been selected, as its works similarly to what has been depicted in the overall project scenario (see section 7) transfering data into Oracle using the object map file.

Here we give a detailed description of the aspects of XML-DBMS that it is necessary to know to accomplish the most common tasks.

### 9.1 What is XML-DBMS?

By definition, XML-DBMS is a set of Java packages that are meant for providing and implementing mapping rules between objects embedded in the Object XML files and tables inside the relational databases.

It preserves the followings

❑ The hierarchical structure of an XML document

❑ Data (character data and attribute values) in that document.

❑ The order in which the children at a given level in the hierarchy appear.

To run XML-DBMS, the following software is needed:

- XML-DBMS, which can be downloaded from:
  http://www.bourret.com/xmldbms/index.htm

- JDK (Java Development Kit) 1.1.x or 1.2.x, which can be downloaded from:
  http://java.sun.com/products/jdk/1.1/index.html

- The Oracle database, (tests have been carried out with Oracle8i version 8.0.6 and 8.0.7 and Oracle9i)

**An XML parser is required for XML-DBMS**: XML parsers are available from many companies, such as Oracle, Sun, IBM, and Microsoft (DataChannel). In addition, Open Source parsers are available from many organizations and individuals, such as James Clark [17], OpenXML [18], and Apache [19]. For the use cases reported in this paper the Oracle XML parser version 2 has been used.

## 9.2 How XML-DBMS views an XML document:

XML-DBMS views an XML document as a tree of objects and then uses an object-relational mapping to map these objects to a relational database.

In this view,
- Element types generally correspond to classes and attributes and PCDATA correspond to properties.
- Child element types are generally viewed as pointed-to classes; that is, an interclass relationship exists between the classes corresponding to parent and child element types.

## 9.3 Mapping XML Documents to the Database

The user specifies how element types, attributes, and PCDATA are viewed, as well as how to map this view to the database. This information is contained in a Map object, which is created by a map factory.

## 9.4 The XML-DBMS Mapping Language

The XML-DBMS mapping language is a simple; it is an XML-based language that describes both how to construct an object view for an XML document and how to map this view to a relational schema. For complete information about the XML-DBMS mapping language, please follow the web link of "mapping language DTD" [11].

## 9.5 Table and Column Names

Table and column names in the map document must **exactly match** the names stored in the database. This happens because Oracle converts table and column names in a CREATE TABLE statement to all upper or all lower case.

For example, suppose the name SALES has been used in a CREATE TABLE statement. The database might store this name as SALES, in that case the name SALES must be used in the map file.

## 9.6 Transferring Data between XML Documents and the Database

XML-DBMS has two classes for transferring data between XML documents and the database: DOMToDBMS transfers data from XML documents to the database and DBMSToDOM transfers data in the opposite direction. Both classes use DOM trees as intermediate forms of the XML document (SAXToDBMS and DBMSToSAX classes are planned for a future release of XML-DBMS. These should help solve some of the scalability problems encountered by using DOM trees.)

**Transferring Data to the Database:** while transferring data the DOMToDBMS class transfers that data from a DOM tree to the database according to a given Map.

**Key (Object ID) Generators:** KeyGeneratorImpl generates unique 4-byte integers based on a value stored in a special table. Before using KeyGeneratorImpl, following should be done:

- A table will be created named XMLDBMSKey with a single INTEGER column named HighKey.
- A single row to this table with HighKey set to 0.

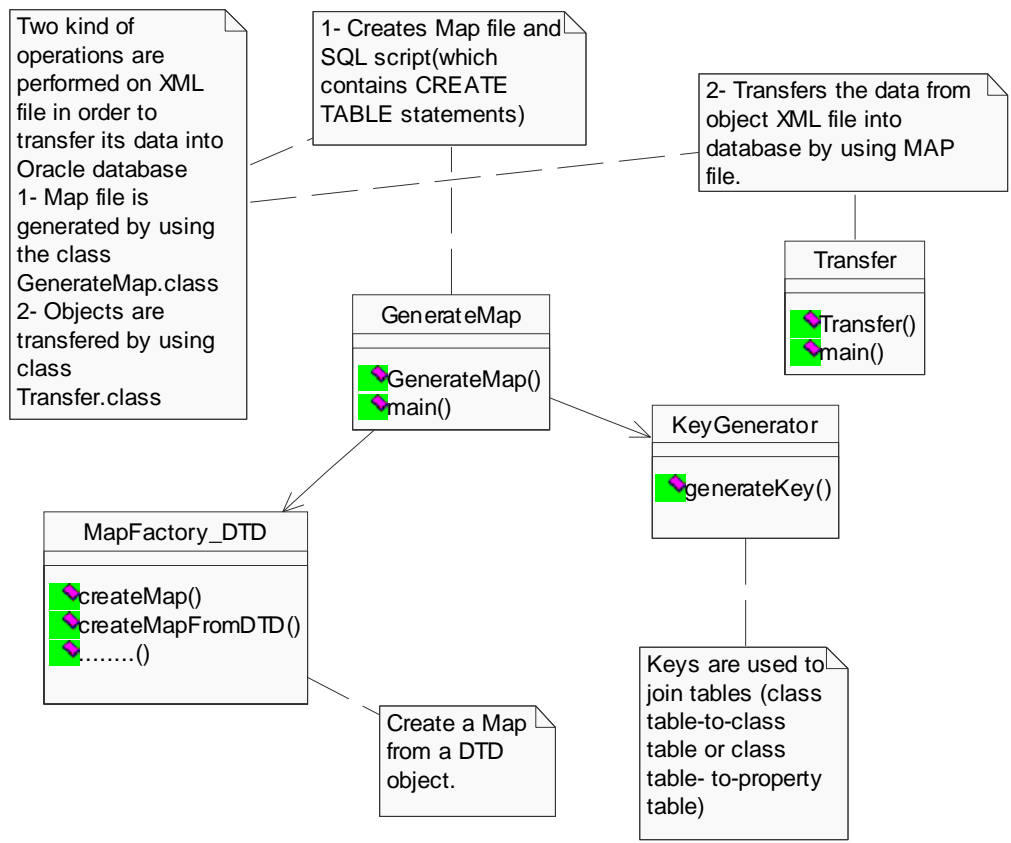

**Fig. 9**: shows the higher-level class structure that how data is transformed into Relational database details is are as follows.

## 9.7 GenerateMap

GenerateMap is java class of XML-DBMS that generates a map and a set of CREATE TABLE statements from a DTD, an XML document containing or referring to a DTD, or a DDML schema document. The map is saved in a file with the **.map** extension and the CREATE TABLE statements are saved in a file with the **.sql** extension. It shows how to use the MapFactory_DTD and Map classes.

To run GenerateMap the following command is used:

     *Java GenerateMap <DTD or XML document>*

For example, to generate a map from structure information of the objects residing in the Objectivity/DB, use following command:

// CRISTALSCh.xml is DTD based schema file of CRISTAL DB.

     *Java GenerateMap CRISTALSCh.xml*

The GenerateMap application requires an ODBC data source named "xmldbms" and an ODBC driver for that database. It does not require that the database contain any tables -- it simply needs to retrieve information from the database about how to construct the CREATE TABLE statements.

This application is using Oracle ver.2 XML parser. However, the Xerces and Sun XML parsers can also be used.

## 9.8 MapFactory_DTD

MapFactory_DTD is a Java class that is used to create a Map from a DTD object. MapFactory_DTD constructs tables and columns in which the element types and attributes described in the DTD object can be stored, and then it creates a Map that maps the element types and attributes to these tables and columns. The resulting Map cannot be used immediately with DBMSToDOM or DOMToDBMS because it is required to set a connection with the database. Furthermore, it is possible that the tables referred to by the map don't yet exist. However, it can be serialized as a mapping document or used to generate CREATE TABLE statements.

For example, the following code creates a map from the DTD document.dtd, creates the tables, sets the connection, and then transfers data to the database:

        // Instantiate a new map factory and create a map.

        factory = new MapFactory_DTD();
        map = factory.createMapFromDTD(src, MapFactory_DTD.DTD_EXTERNAL, true,   null);

/ * Create the tables used by the map. Note that this function calls Map.getCreateTableStrings(), then executes each string in a JDBC Statement. */

        CreateTables(map);

        // Set the database connection, and then transfer the data to the database.
        map.setConnection(conn);
        domToDBMS = new DOMToDBMS(map);
        domToDBMS.storeDocument(doc);

MapFactory_DTD constructs tables and columns. In order to explain here that these tables and columns are not actually created in the database; to do this, the current application must retrieve CREATE TABLE strings from the resulting Map and execute them in JDBC statements. The reason for this is that applications will commonly want to change the table structure predicted by MapFactory_DTD before actually creating tables or simply use this factory as a tool for creating Maps, which can be serialized with the Map.serialize() method. SQL statements created by MapFactory_DTD can be run in the Oracle server directly.

   MapFactory_DTD generates SQL statements for creating tables and columns in the following context:

- ❑   For each element type that has attributes or child elements, a table is generated. Then there is a primary key (PK) column, one column for each single-valued attribute, one column for each singly-occurring child element type that contains only PCDATA and has no attributes, an (optional) order column for each child element type column, and one foreign key (FK) column for each parent element type. If the element type has attributes and PCDATA but no child element types, then there is also a column for its PCDATA. Note that the PK column appears only if needed to link to a child table or if the element type is a potential root element type.

- ❑   If an attribute is multi-valued (IDREFS, NMTOKENS, or ENTITIES), it is stored in a separate table, with an element type FK column, a value column, and an (optional) order column.

- ❑   If a child element type that contains only PCDATA and has no attributes can occur multiple times in its parent, it is stored in a separate table, with a parent element type FK column, a value column, and an (optional) order column.

- ❑   Except as noted above, PCDATA is stored in a separate table with an element type FK column,

a value column, and an (optional) order column.

❑ The code also guesses at what the legal root element types are. An element type is considered to be a root if it has no parents. If all element types have parents, then all element types are made legal roots.

## 9.9 Transfer

**Transfer** is a Java application and it is also a part of XML-DBMS and has an important role in data transfer. This application accepts a map document name, an XML document name, and (in the case of transferring data from the database to an XML document) a table name. It transfers data in the specified direction. It shows how to use the MapFactory_MapDocument, DOMToDBMS, DBMSToDOM, and Map classes.

When transferring data from an XML document to the database, the following command can be used:

Java Transfer -todbms CRISTALSCh.map ObjectTest.xml

Where CRISTALSCh.map is map-file and ObjectTest.xml is an xml-file that contains actual data. Before using the Transfer, the following things should be done earlier:

❑ An ODBC data source will be created with the name of "xmldbms" and that will points towards database.
❑ Tables will be created in the database in which data will be inserted
❑ A table will be created and then initialized with the name of XMLDBMSKey.

The application "Transfer" is also hard-coded.

# 10. Example of Object mapping in Oracle:

Listed example shows the transformation of object in Oracle DB, for the understanding of reader this example is showing transformation of a single object.

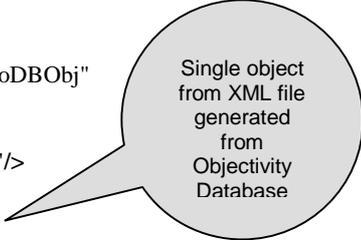

```
<Object id="3-3-3-17" typename="ooMapElem" typnumber="4002">
    <Database id="3-0-0-0" name="PRESHOWER_Config" typename="ooDBObj"
    typnumber="1004"/>
    <Container id="3-3-3-1" name="TEST_Cont" typename="ooContObj"/>
    <Attributes>
    <String name="_key" type="ooVString">
      <StringElement index="0">2</StringElement>
    </String>
<Ref name="_value" referencedClass="ooRef(ooObj)"><RefElement index="0">3-3-3-15</RefElement></Ref>

<Ref name="_map" referencedClass="ooShortRef(ooMap)"><RefElement index="0">3-3-3-2</RefElement></Ref>

<Ref name="_next" referencedClass="ooRef(ooMapElem)"><RefElement index="0">3-3-3-74</RefElement>
    </Ref>
    </Attributes>
</Object>
```

Single object from XML file generated from Objectivity Database

**Table Object**

| TYPENAME | OBJECTPK* | ID |
|---|---|---|
| OoMapElem | 2488 | 3-3-3-17 |

**Table Database**

| NAME | TYPENAME | ID | OBJECTFK* |
|---|---|---|---|
| PRESHOWER_Config | ooDBOb | 3-0-0-0 | 2488 |

**Table Container**

| NAME | TYPENAME | ID | ObjectFK* |
|---|---|---|---|
| TEST_Cont | ooContObj | 3-3-3-1 | 2488 |

**Table Attributes**

| ATTRIBUTEPK* | OBJECTFK* |
|---|---|
| 2489 | 2488 |

**Table String**

| STRINGPK* | NAME | TYPE | ATTRIBUTEFK* |
|---|---|---|---|
| 2493 | "_key | ooVString | 2489 |

**Table Ref**

| NAME | REFERENCEDCLASS | REFPK* | ATTRIBUTESFK* |
|---|---|---|---|
| "_next | ooRef(ooMapElem) | 2490 | 2489 |

**Figure shows:** Object 3-3-3-17 transferred into ORACLE9i DB.

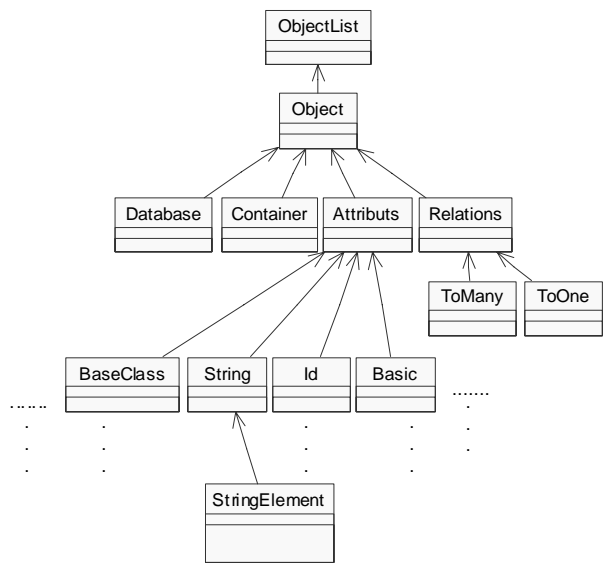

The data of Object 3-3-3-17 is in this Class hierarchy

## 11. Prototype

A test bed is provided with this solution (Migration of objects in Oracle database). As a first step (1) we have two files generated from objectivity database, i.e. Schema XML and Object XML file, the data in the Object XML is to be transferred into Oracle database. To transfer this data two kinds of files are required (2). First one is the SQL script file, that will be used to create tables into oracle database, and second one is Map file, that will contain the information for mapping of objects into Oracle database table columns.

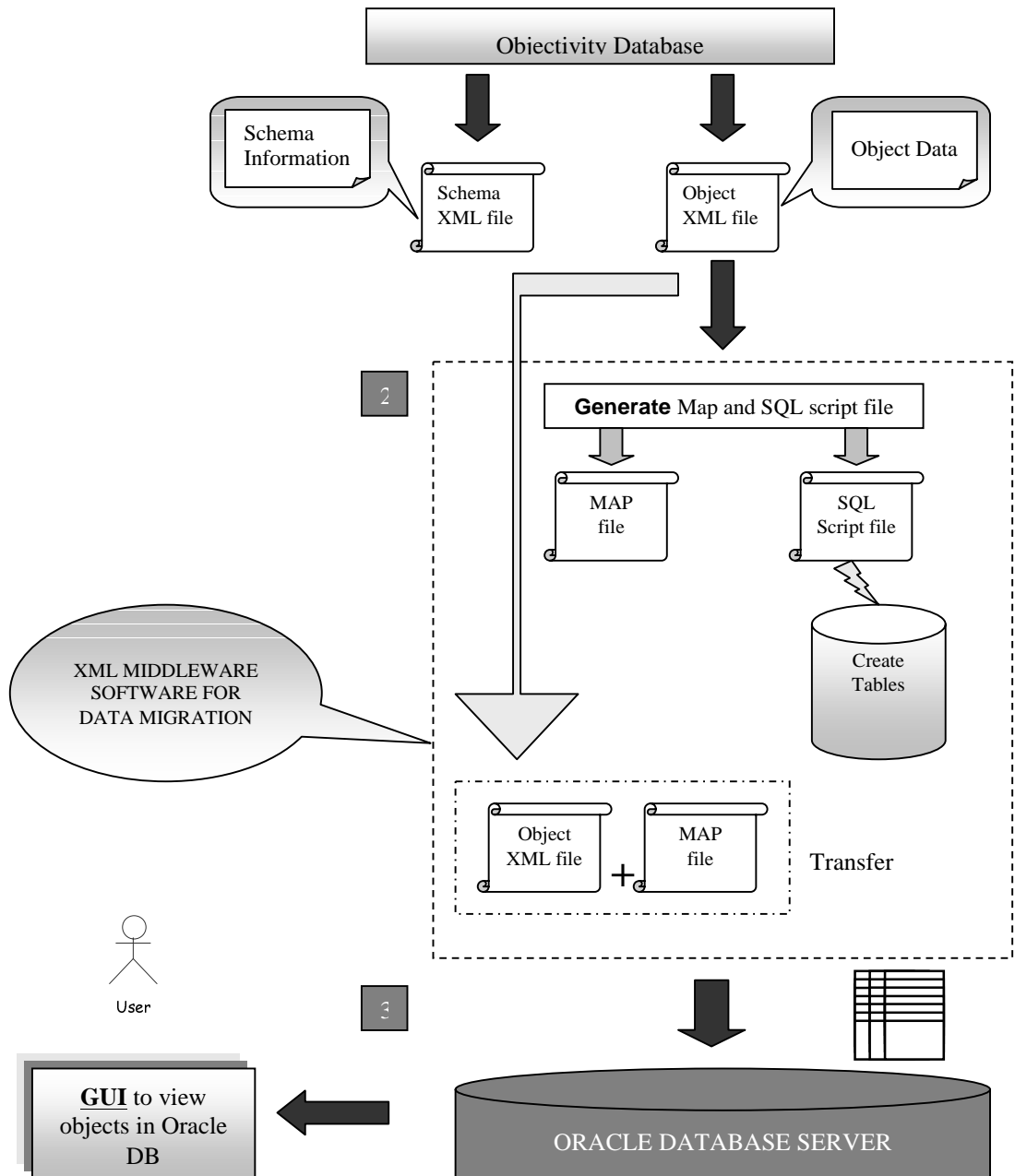

After creating the tables in Oracle the objects in Object XML file will be transferred into oracle tables (generated by the SQL create table statements from SQL script file). Moreover we also provided a graphical user interface (GUI) to view the migrated objects into Oracle database. The GUI will also help to see the primary foreign key relationship between objects as described in the example (Object mapping in Oracle).

We have tested our proposed solution by transferring an Object XML file into Oracle database. File size was 2.64 MB, and the transfer time was 180 seconds on P-III computer.

## 12. Future

At the time when this paper has been submitted XML-DBMS version 1.01 is stable. However, much work is going into XML-DBMS version 2.0, which will have increased functionality. XML-DBMS 2.0 and 3.0 [10], will have new features such as storing data in multiple databases, Character Encoding, Multiple key generators, Mapping multi-valued properties to the parent table etc.

At present a schema file that is DTD-based has been used (to generate schema information and map file). In future XML schema can be used for that function. Currently XML-DBMS is providing APIs for DOMToDBMS, DBMSToDOM, where as SAXToDBMS and DBMSToSAX classes are planned for a future release of XML-DBMS. SAXToDBMS and DBMSToSAX are supposed to be significantly faster than DOMToDBMS and DBMSToDOM as long as *order* is not important

The biggest changes in 2.0 will be *'changes to Map objects'* this will make XML-DBMS more efficient, writing new map
factories will be easier, and it will be possible to efficiently recombine map fragments.

One feature of object-oriented databases is that they assign object IDs. XML-DBMS does not currently assign object IDs to leaf nodes, this is the area that needs to be explored as a future study.

# 13. Conclusion

In this paper a mechanism to handle the mapping of objects into relational tables and modeling the concepts of object oriented programming into relational table structures are investigated. Different XML middleware products have been tested (including both freeware and commercial products) available for mapping objects to tables. Various test were conducted to discover the strengths and limitations of these products and based on the results the products (as mentioned above) have been selected, which were close to the project's design and needs. This was actually done to analyze the structure of the objects residing in the objectivity database (at source) and to extract the underlying metadata for converting it in a format suitable for designing schema in the RDBMS i.e. Oracle. Finally, the task was to pull out the objects from the object XML file generated from Objectivity database (at source) to migrate them in the Oracle database by making use of the map file.

# 14. Acknowledgements

CERN support in providing data-files for testing the software and verifying the results is acknowledged in the current research work.

# Appendix

A. Software requirements / Run pre configured Test Bed
B. XMLDBMS architecture model
C. CPU and memory state while transformation of object data into Oracle9i database.
D. Glossary of Terms:
_______________________________________________________________________________

# Appendix A

## Software Requirements:

- **Oracle8i/9i database server:**
  Install Oracle8i/9i database server version 8.1.7, 9i or higher.

- **JDK (Java Development Kit) 1.1.x or 1.2.x:**
  Download: http://java.sun.com/products/jdk/1.1/index.html

- **An XML parser written in Java:**
  XML parsers are available from many companies, such as Oracle, Sun, IBM, and Microsoft (DataChannel). In addition, Open Source parsers are available from many organizations and individuals, such as James Clark, OpenXML, and Apache.

- **JDBC driver for your database:**
  Most relational databases are shipped with JDBC and ODBC drivers. If you have an ODBC driver but not a JDBC driver, you can use an JDBC-ODBC bridge as your JDBC driver. This converts JDBC calls to ODBC calls. An experimental (and therefore somewhat buggy) JDBC-ODBC bridge is shipped with the JDK; JDBC-ODBC bridges are available from other companies as well. Note that the quality of JDBC drivers varies considerably, so if one JDBC driver does not work, you can try another.

# CONFIGURATION TO RUN TEST BED

Steps to use the pre configured test bed that transfers data from XML file into Oracle database.

We have tested it with a XML file generate form CRISTAL database on windows2000 professional and server, for the configuration of that transformation and middleware transformation follow there steps.

## Installation Issues

### Pre requirements:

- Oracle8i/9i database server version 8.1.7, 9i or higher

- JDK (Java Development Kit) 1.1.x, 1.2.x or higher
  Download JDK: http://java.sun.com/products/jdk/1.1/index.html

- I order to run GUI provided with this test bed to see the transferred objects in Oracle database you need to install Oracle Forms (*Run Time*) Developer 6i

## Steps to run pre configured Test Bed

**1.** After completing the pre requirements (configuration of oracle and JDK), unzip the provided install.zip file and simply place two folders (xmldbms and xerces-J) and files "xmlparserv2.jar" (which is a XML parser) and other files in JDK folder.

**2.** Now add these system variables (go in the properties of MyComputer ➔ advanced ➔ system variable and add a variable with the name of "classpath" then add following values in it

C:\jdk1.3\xerces-J\xerces.jar;

C:\jdk1.3\xerces-J\xercesSamples.jar;

C:\jdk1.3\xmldbms\xmldbms_jar;

C:\jdk1.3\xmlparserv2.jar

**3.** After completing Oracle user configuration, now enter user name and password in ODBC Driver setup (Microsoft ODBC Driver for ORACLE) and enter data source name "xmldbms". (See fig. on next page)

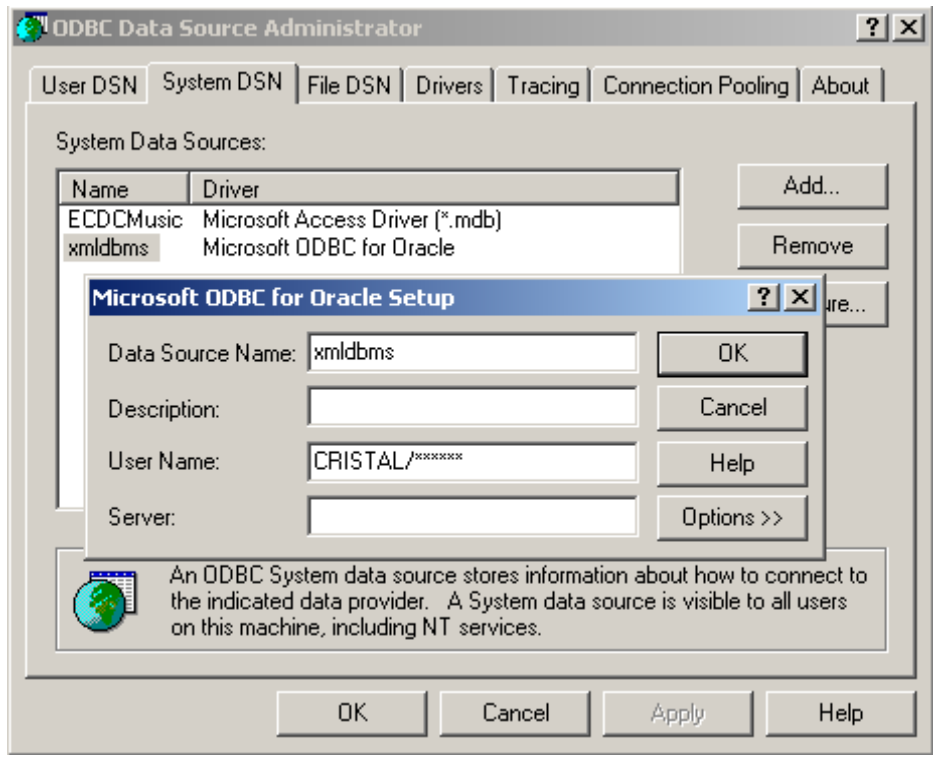

Here user name is CRISTAL and password is *****

**4.** In order to create tables in Oracle database run sql script "CreateTables.sql" provided with this bed or simply cut and paste all the text provided in the CreateTables.sql file in your ORACLE SQL prompt.

**5.** Now go into the folder xmldbms➔ Samples and run bath file TransferData.bat file. The ObjectTest.xml file will be transferd by using ObjectTest.map file into Oracle database.

## GUI to see the migrated data in Oracle:

After the transformation the data in Oracle can be viewed form SQL prompt or by using the GUI, provided with this test bed. To run it just simply place the CRISRAL-GUI.fmx file any where or on c:\ and double click on it (then enter user name and passward same as given in ODBC setup). All the information of using the trouble-free GUI can be seen by its help, tool tips etc. If the CRISRAL-GUI.fmx will give any error then you can use CRISRAL-GUI.fmb and run it from Oracle form developer6i.

# Appendix B

**XML-DBMS Architecture Model:**

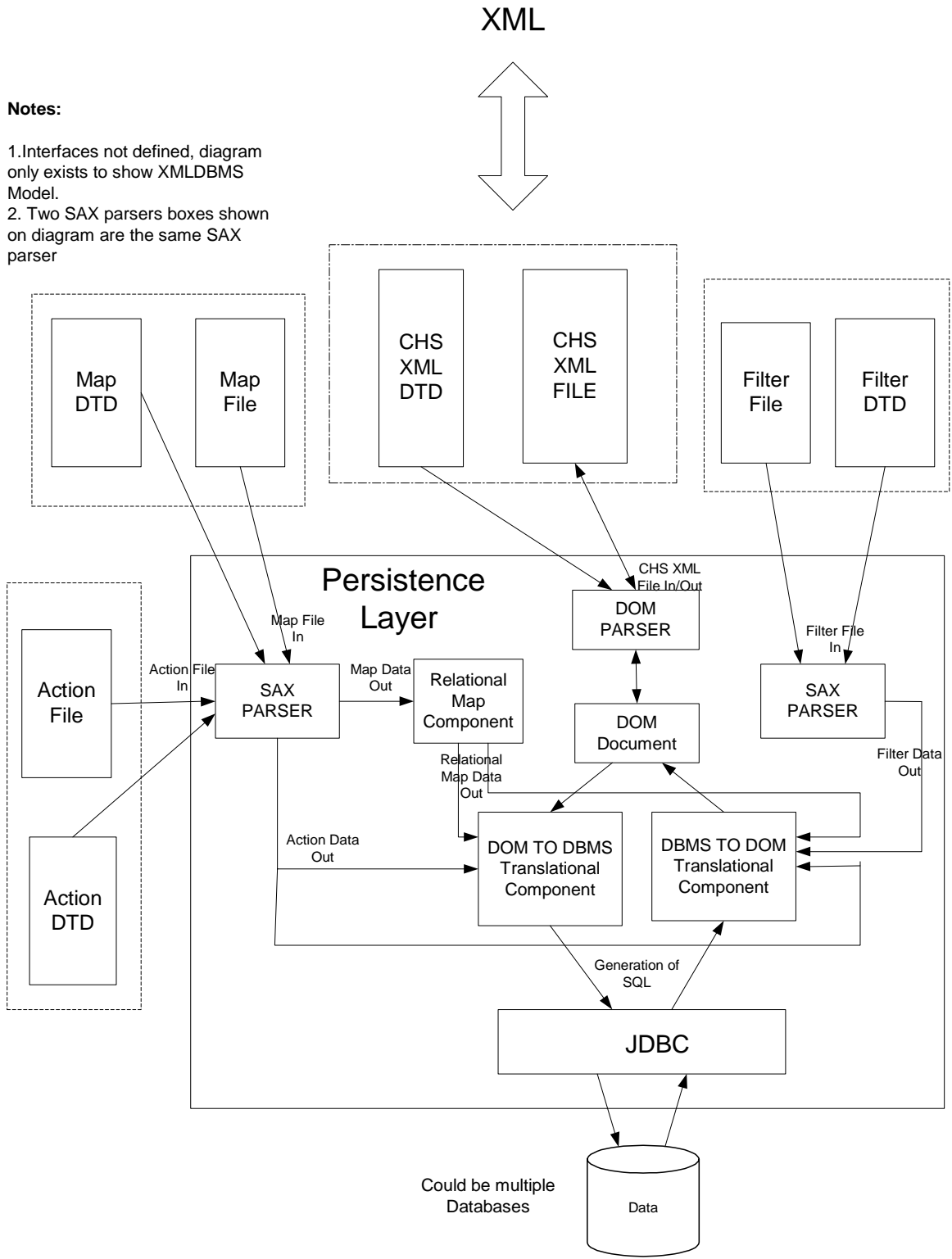

Source: XML-DBMS Group [10]

# Appendix C

**CPU and Memory state before transformation of Objects into Oracle9i database Server.**

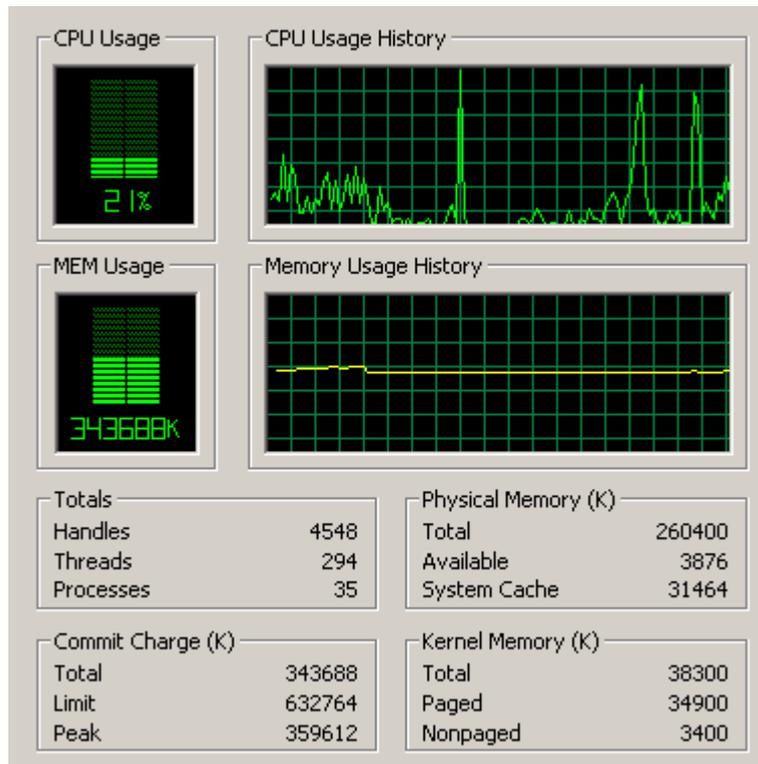

**CPU and Memory state at the time of Transformation of Objects into Oracle9i database Server.**

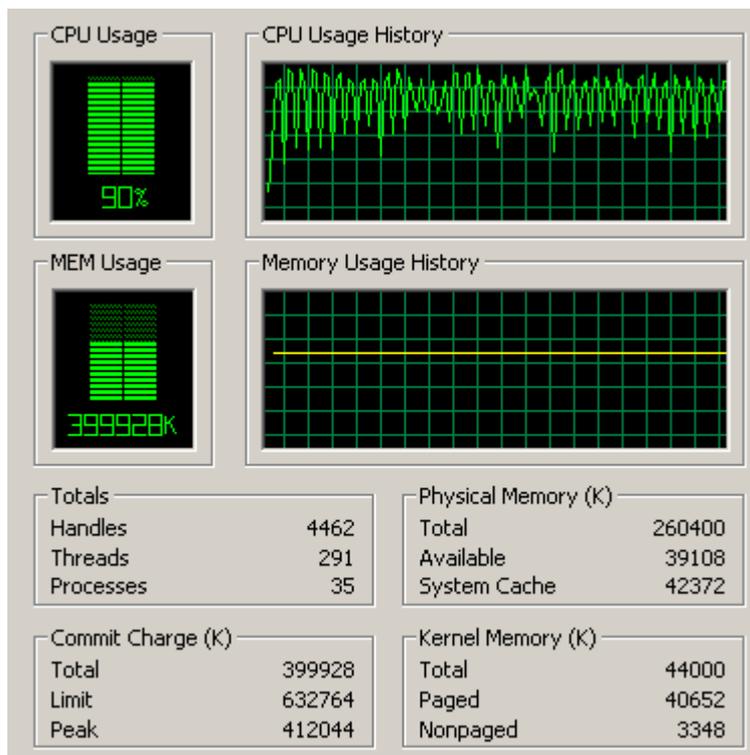

# Appendix D

## Glossary of Terms:

This section describes the key terms that I have used throughout this document.

**Aggregation:** Represents "is-part-of" relationships.

**Association:** Relationships, associations, exist between objects. For example, customers BUY products.

**Associative table:** A table in a relational database that is used to maintain a relationship between two or more other tables. Associative tables are typically used to resolve many-to-many relationships.

**Client:** A single-user PC or workstation that provides presentation services and appropriate computing, connectivity, and interfaces relevant to the business need. A client is also commonly referred to as a "frontend."

**Client/server (C/S) architecture:** A computing environment that satisfies the business need by appropriately allocating the application processing between the client and the server processes.

**Data dictionary:** A repository of information about the layout of a database, the layout of a flat file, the layout of a class, and any mappings among the three.

**Database server:** A server which has a database installed on it.

**Distributed objects:** An object-oriented architecture in which objects running in separate memory spaces (i.e. different computers) interact with one another transparently.

**Key:** One or more columns in a relational data table that when combined form a unique identifier for each record in the table.

**Object identifiers (OIDs):** A unique identifier assigned to objects, typically a large integer number. OIDs are the object-oriented equivalent of keys in the relational world.

**Pattern:** The description of a general solution to a common problem or issue from which a detailed solution to a specific problem may be determined. Software development patterns come in many flavors, including but not limited to analysis patterns, design patterns, and process patterns.

**Persistence:** The issue of how to store objects to permanent storage. Objects need to be persistent if they are to be available to you and/or to others the next time your application is run.

**Persistence classes:** Persistence classes provide the ability to permanently store objects. By encapsulating the storage and retrieval of objects via persistence classes you are able to use various storage technologies interchangeably without affecting your applications.

**SQL:** Structured Query Language, a standard mechanism used to CRUD records in a relational database.

**SQL statement:** A piece of SQL code.

**Transaction:** A transaction is a single unit of work performed in a persistence mechanism. A transaction may be one or more updates to a persistence mechanism, one or more reads, one or more deletes, or any combination thereof.